\documentclass[a4paper,fleqn]{cas-sc}

\usepackage[numbers]{natbib}
\usepackage{mathtools}
\usepackage{eurosym}
\usepackage{subcaption}
\usepackage[scientific-notation=true]{siunitx}

\usepackage{framed} 
\usepackage{multicol} 

\makeatletter
\def\@seccntformat#1{\@ifundefined{#1@cntformat}%
   {\csname the#1\endcsname\quad}
   {\csname #1@cntformat\endcsname}
}
\makeatother
 
\usepackage{nomencl} 
\usepackage[boxruled]{algorithm2e} 
\makenomenclature
\renewcommand*\nompreamble{\begin{multicols}{2}}
\renewcommand*\nompostamble{\end{multicols}}

\renewcommand\nomgroup[1]{%
  \item[\bfseries
  \ifstrequal{#1}{A}{Abbreviations}{%
  \ifstrequal{#1}{V}{Variables}{%
  \ifstrequal{#1}{P}{Parameter}}}%
]}

\begin{document}
\let\WriteBookmarks\relax
\def\floatpagepagefraction{1}
\def\textpagefraction{.001}
\shorttitle{Adequacy of time-series reduction}
\shortauthors{Göke}

\title [mode = title]{Adequacy of time-series reduction for renewable energy systems}

\author[1]{Leonard Göke}[type=editor,auid=000,bioid=1]
\cormark[1]
\cortext[cor1]{Corresponding author.}
 \ead{lgo@wip.tu-berlin.de}
\author[1]{Mario Kendziorski}[auid=000,bioid=1]

\address[1]{Berlin University of Technology, Workgroup for Infrastructure Policy (WIP), Berlin, Germany.}

\begin{abstract}
To reduce computational complexity, macro-energy system models commonly implement reduced time-series data. For renewable energy systems dependent on seasonal storage and characterized by intermittent renewables, like wind and solar, adequacy of time-series reduction is in question. Using a capacity expansion model, we evaluate different methods for creating and implementing reduced time-series regarding loss of load and system costs.

Results show that adequacy greatly depends on the length of the reduced time-series and how it is implemented into the model. Implementation as a chronological sequence with re-scaled time-steps prevents loss of load best but imposes a positive bias on seasonal storage resulting in an overestimation of system costs. Compared to chronological sequences, grouped periods require more time so solve for the same number of time-steps, because the approach requires additional variables and constraints. Overall, results suggest further efforts to improve time-series reduction and other methods for reducing computational complexity.
\end{abstract}

\begin{highlights}
\item Dependence on renewables and storage reduces the adequacy of time-series reduction.
\item Implementation as a re-scaled sequence prevents loss of load, but overestimates costs.
\item Further efforts to improve time-series reduction or other methods are needed.
\end{highlights}


\maketitle

\begin{table*}[pos=!t]
   \begin{framed}
     \printnomenclature[1cm]
   \end{framed}
\end{table*}

\nomenclature[A, 01]{TSR}{time-series reduction}
\nomenclature[A, 02]{CAES}{compressed air energy storage}

\nomenclature[P, 01]{$\alpha$}{parameter to re-scale reduced time-series}
\nomenclature[P, 02]{$\beta$}{parameter to correct compression of redcued time-series}
\nomenclature[P, 03]{$dem_{t}$}{demand in period $t$}
\nomenclature[P, 04]{$cf_{t}$}{capacity factor in period $t$}
\nomenclature[P, 05]{$invC_{i}$}{specific investment costs}
\nomenclature[P, 06]{$varC$}{specific variable costs}
\nomenclature[P, 07]{$m$}{length of the reduced time-series in hours}

\nomenclature[V, 01]{$InvC$}{total investment costs}
\nomenclature[V, 02]{$VarC$}{total variable costs}
\nomenclature[V, 03]{$Gen_{t}$}{generation in period $t$}
\nomenclature[V, 04]{$St_{t}^{out/in}$}{storage in- and output in period $t$}
\nomenclature[V, 05]{$St_{t}^{size}$}{storage level in period $t$}
\nomenclature[V, 06]{$Capa_{gen}$}{power capacity of generation}
\nomenclature[V, 06]{$Capa_{st}$}{power capacity of storage in- and output}
\nomenclature[V, 07]{$Capa_{size}$}{energy capacity of storage}

\section{Introduction} \label{1}

Mitigation of climate change and decarbonization of the energy system requires replacing fossil fuels with renewable energy, especially wind and solar. To emphasize how many research questions related to decarbonization call for an extensive temporal, spatial, and sectoral scope, \citet{Levi2019} introduce the term \textit{macro-energy systems}. A key tool for analyzing how renewable energy can replace fossil fuels in macro-energy systems are long-term planning models. These models can be deployed to compute comprehensive decarbonization scenarios or study more specific questions, like trade-offs when deploying technologies or options for placing renewables \citep{Auer2020,Neumann2020,Trondle2020}.

Due to their extensive scope, long-term planning models have to rely on special methods to keep their computational complexity manageable. The most common approach is to use a reduced time-series, that is ought to preserve all key characteristics of the original time-series, but contains fewer elements, aiming for a favorable trade-off between complexity and accuracy \citep{Hoffmann2020,Buchholz2019}.

The characteristics of energy systems with high shares of renewables put the use of reduced time-series into question. Unlike thermal generation, supply from wind and solar fluctuates and requires energy storage to match generation with demand. While moderate shares of renewables only require comparatively small amounts of short-term storage, e.g. battery or pumped-hydro, needs change substantially as shares further increase towards 100\%. In this case, in addition to short-term storage, systems increasingly depend on long-term storage, for example based on synthetic hydrogen, to balance supply and demand across seasons \citep{Zerrahn2017, Schill2020, Jenkins2018}. If reduced time-series cannot adequately capture fluctuations and storage requirements, models will determine results, like installed capacities, that are sub-optimal or cannot fully satisfy demand. 

Although time-series reduction (TSR, also referred to as time-series aggregation) is often applied to determine capacities at high shares of renewables, evaluations of its accuracy for such systems are rare. When novel methods of TSR are introduced, their accuracy is typically only tested for existing systems with moderate shares of renewables \citep{Teichgraeber2019a,Nahmmacher2016,Almaimouni2018,Poncelet2016b}.

The highest renewable share evaluated in the existing literature is 90\% \citep{Pfenninger2017}. The analysis compares resulting capacities and system costs for a hourly resolution of 8,760 time-steps to various reduced time-series with 168 to 2,920 time-steps derived by down-sampling, heuristics, k-means and hierarchical clustering. Results show that the accuracy of TSR greatly depends on the reduction method and length of the reduced time-series, but generally decreases at higher shares of renewables. In addition, the inclusion of short-term storage is found to increase accuracy of TSR, whereas the impact of seasonal storage is not investigated.

\citet{Mallapragada2018} analyze the impact of TSR on capacity expansion models up to a renewable share of 70\% and without any storage. For this purpose, 16 steps grouped into independent slices are compared to a chronological sequence with 288 steps, both computed using k-means clustering. When testing results with a detailed dispatch model, capacities computed using grouped periods cannot meet between 0.15\% and 0.5 \% of demand. For the chronological sequence unmet demand never exceeds 0.1\% and strongly increases with the renewable share. For both methods, low temporal resolution is found to overestimate solar investments while underestimating wind and gas power plants.  

Similarly, \citet{Reichenberg2018} test TSR at a renewable share of 50\% omitting any storage technologies. With k-means clustering times-series of up to 200 steps are derived and implemented into a capacity expansion model, either using an "integral" method with limited suitability for more complex applications or, analogously to \citeauthor{Mallapragada2018}, grouped periods. Again, a low resolution is found to overestimate solar at the expense of wind and gas.

Instead of enforcing a specific renewable share, \citet{Merrick2016} analyze how adding wind and solar to a stylized model impacts the accuracy of TSR. For this purpose, the paper adopts a heuristic reduction algorithm and find that the reduced time-series needs to be extended from 10 to 1,000 steps for the model to remain accurate when wind and solar are added. 

In addition to the literature cited above, two other types of studies on TSR can be identified. First, there are several studies investigating high shares of renewables and even include storage, but are limited to operation of pre-set capacities \citep{Scott2019, Raventos2020}. At high shares of renewables, operational costs are negligible compared to investment, which renders the studies inapplicable for comparison in the context of this paper. A second group of studies considers TSR and capacity expansion, but does not analyze macro-energy systems, which also hinders comparison. Instead the focus is on models of small-scale systems, like residential homes or industrial sites, that have different characteristics, for example the option of flexible supply from the grid and modeling specific units requires integer constraints \citep{Fahy2019,Schuetz2018}.

In summary, previous work finds that increasing shares of wind and solar reduce the accuracy of TSR in large-scale capacity expansion models, though adequate reduction methods and increased temporal resolution mitigate this effect. So far evaluations are limited to renewable shares up to 90 \% and consequently also neglect the need for seasonal storage arising if shares increase further. Therefore, this paper evaluates the adequacy of TSR in capacity expansion models of fully renewable macro-energy systems. For this purpose, capacities computed with a reduced time-series are tested with regard to system costs and generation adequacy using a full time-series again. Here, "full" refers to hourly data, which is the standard generally considered adequate for large-scale models. Since these models cover extensive geographical regions, any sub-hourly fluctuations are assumed to be balance out within each region \citep{Brown2018b}.

Our analysis of TSR makes a strict distinction between how reduced time-series are implemented into models and how they are derived from a full time-series. Section \ref{2} and \ref{3} respectively introduce the different methods considered for implementation and derivation of reduced time-series. Section \ref{4} briefly introduces the capacity expansion model used to evaluate the reduced time-series followed by section \ref{5} that discusses the corresponding results. The paper is closed of by a summary of key findings and their implications for future modeling work.

\section{Implementation of reduced time-series} \label{2}

Since literature shows that the way models implement reduced time-series greatly affects results, this paper evaluates TSR for different implementation methods. We strictly separate implementation of reduced time-series from their derivation that is discussed in the following section. This section introduces two fundamentally different approaches to implement reduced time-series and discusses their capabilities to account for storage. 

\subsection{Grouped periods} \label{21}

The first method groups the elements of the reduced time-series into periods and then considers each of these periods separately. In this paper, we will refer to this method as "grouped periods". Historically, this method originates from long-established models for planning energy systems like TIMES or MARKAL \citep{Loulou2016,Kannan2011}. 

There are various terms for "grouped periods" in the literature, that we've decided not to fall back on, because their use is highly inconsistent and ambiguous. For example, some studies use the term "representative days" for this implementation method, while others use the same term in a literal sense referring to days representative for a whole year regardless of how they are implemented \citep{Reichenberg2018,Mallapragada2018}. On the other hand, some studies refer to grouped periods as "time-slices", but others use the same term to describe a method for deriving reduced time-series, not for their implementation into models \citep{Mallapragada2018,Pfenninger2017}.

\begin{figure}
	\centering
		\includegraphics[scale=.4]{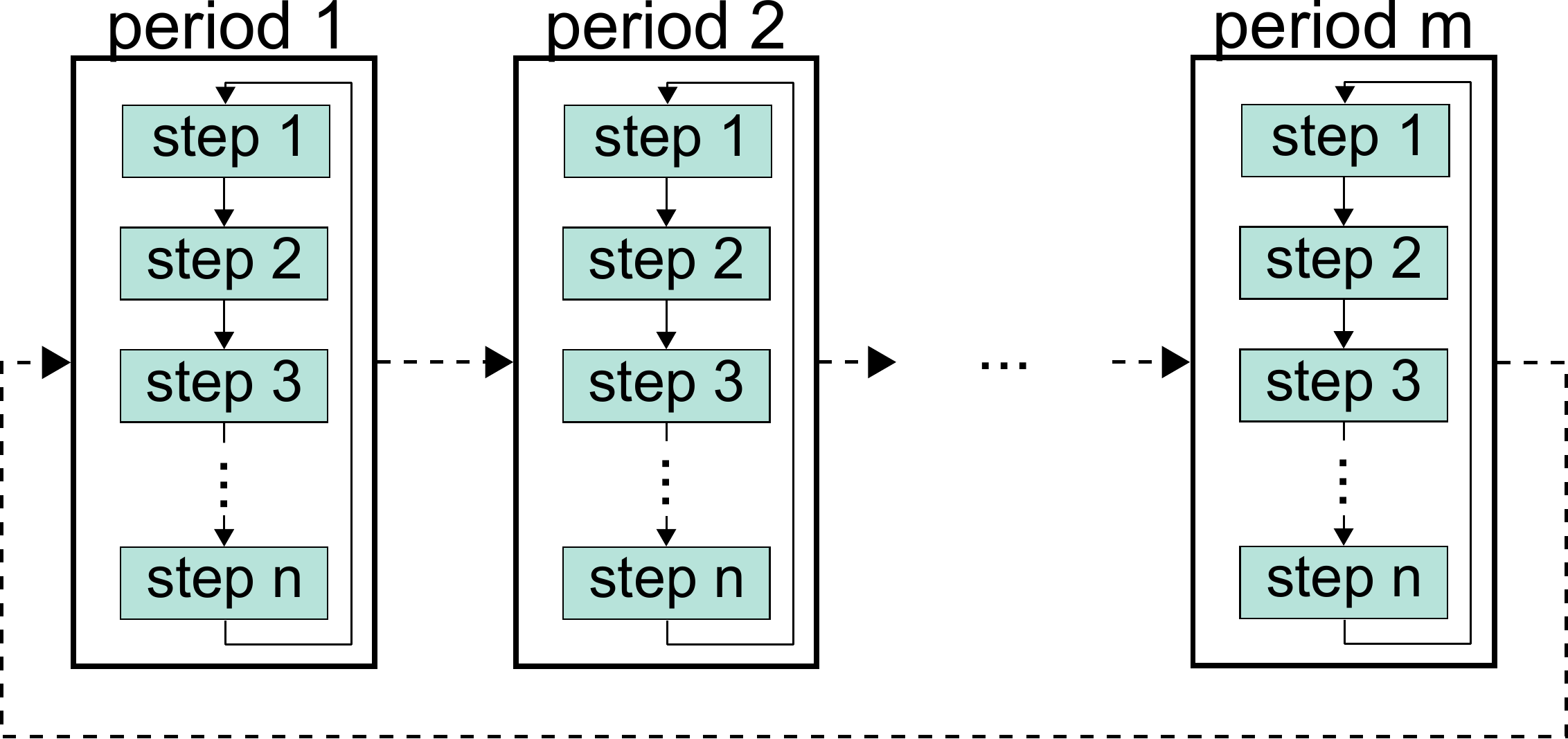}
	\caption{Concept of grouped periods}
	\label{fig:1}
\end{figure}

Fig. \ref{fig:1} illustrates the method: A reduced time-series is grouped into several independent periods each containing a chronological sequence of time-steps. The length of these sequences is arbitrary, although most commonly each period is set to represent one day. Since time-steps from different periods are considered independently, computational effort is reduced, and periods can be weighted to reflect characteristics of the original time-series more closely.

However, the use of independent periods faces limitations when modeling storage. As indicated by the arrows in Fig. \ref{fig:1}, cyclic conditions connecting the storage levels of consecutive time-steps can only be enforced within each period. This prevents the exchange of energy across periods, for example charging in period 1 and discharging in period 2. While this simplification can argued to be reasonable for short-term storage, it cannot model seasonal storage---a key component of renewable energy systems as described in section \ref{1} \citep{Zerrahn2017, Schill2020, Jenkins2018}. Several studies identified and addressed this drawback by extending the original approach that is limited to short-term storage to include season storage as well \citep{Kotzur2018a,Arango2018}. 

For our test on the accuracy of TSR in renewable energy systems, we will apply the method introduced in \citet{Kotzur2018b}, because it is most widely used for large-scale capacity expansion problems. Instead of enforcing an independent cyclic condition for each period, \citeauthor{Kotzur2018b} introduce a new variable for the storage levels across periods, which is illustrated by the dotted line in Fig. \ref{fig:1}. This intra-period variable is subject to a yearly cyclic condition replacing cyclic conditions in each grouped period. Its value is computed from the net-balances of charging and discharging within each, formerly independent, period of the reduced time-series. For example, consider a grouped period from the reduced time-series, that was selected to represent 12 periods of the original time-series and consequently has a weight of 12. If charging and discharging of a storage technology in the grouped period nets to 5 GWh, 5 times 12 GWh will consecutively be added to the intra-period storage level. In addition, the capacity limits on each inter-period storage level are replaced with limits on the sum of the intra-period variable and net-charging at each time-step. In summary, instead of enforcing the same storage levels for each grouped period, the approach enforces charging and discharging, the first derivative of the storage level, to be the same for each grouped period.

\subsection{Chronological sequence} \label{22}

\begin{figure}
	\centering
		\includegraphics[scale=.4]{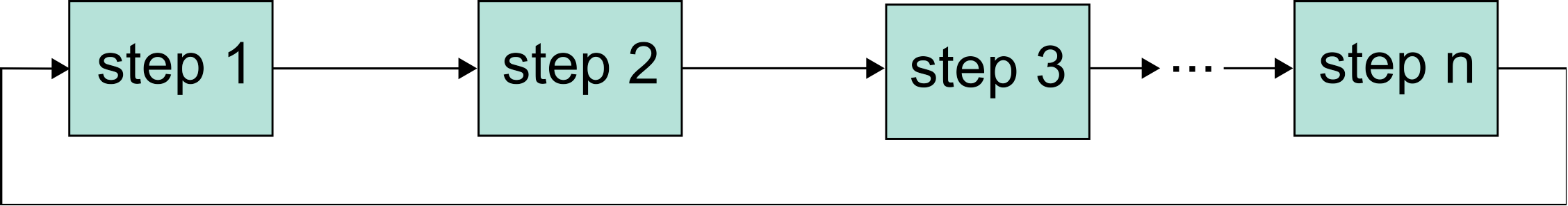}
	\caption{Concept of chronological sequence}
	\label{fig:2}
\end{figure}

Alternatively to grouping periods, a reduced time-series can be implemented as one chronological sequence. This method, illustrated by Fig. \ref{fig:2}, simply puts the time-steps of the reduced time-series into chronological order and connects the first and last step with a cyclic condition for storage, which is analogous to how models represent time-series data that has not been reduced. In contrast to grouped periods, chronological sequences prohibit to assign individual weights to the steps of the reduced time-series. Most commonly, this method is applied by capacity expansion models focused on the power sector \citep{Trondle2020,Neumann2020}.

When implementing a reduced time series as a chronological sequence, there are two ways to achieve consistency with a full time-series and still represent an entire year: First, the size of each time-step can be increased, 2-hour instead of hourly steps for instance will reduce a time-series from 8,760 to 4,380 steps. Second, step-size can be kept constant and instead the year compressed to a smaller number of hours. This means, that, for example, 3,380 hourly time-steps now represent an entire year. Finally, both approaches can be combined, for example, compressing the year to 2,190 hourly time-steps and then applying a 2-hour resolution finally resulting in 1,095 time-steps.

Both approaches have not yet been formalized or discussed with regard to their impact on model results, but are frequently deployed in the literature. Often implementation is closely tied to the method for deriving reduced time-series. For example, down-sampling, a common method for TSR, uses a reduced time-series derived by summing successive hours into new times-steps, which increases step-size but keeps the length of the year constant \citep{Hoffmann2020,Neumann2020,Trondle2020}. In contrast, \citet{Gerbaulet2017} derive a reduced time-series by selecting every n-th hour of the full time-series and implement the selected hours without adjusting their size, effectively compressing the year\footnote{The same method is deployed in \citet{Auer2020}.}. However, it is conceivable in both examples to combine the methods for deriving and implementing reduced time-series differently. Down-sampling could average instead of sum successive hours to keep step-size constant and compress the year instead. On the other hand, hours selected by the method from \citeauthor{Gerbaulet2017} could be scaled up and represent an entire year.

To demonstrate how implementation affects model variables and final results, consider the capacity expansion problem in Eqs. \ref{eq:1a} to \ref{eq:1h}. To differentiate them, variables are written in capital and parameters in lower-case letters. According to the energy balance in Eq. \ref{eq:1b}, the sum of generation $Gen_t$, storage input $St^{in}_t$ and storage output $St^{out}_t$ has to match demand given by the parameter $dem_t$ at each time-step $t$. The following storage balance connects storage in- and output with the storage level $St^{size}_t$ at each time-step $t$. Eqs. \ref{eq:1d} to \ref{eq:1f} enforce capacity constraints on storage in- and output, storage levels and generation ensuring production does not exceed the capacity $Capa_{i}$. For generation, capacity constraints include a capacity factor $cf_t$ that specifies the share of capacity available for generation at time-step $t$. Finally, the objective function Eq. \ref{eq:1a} is composed of total investment costs $InvC$ computed from capacities $Capa_i$ and specific investment costs $invC_i$ in Eq. \ref{eq:1g} and total variable costs $VarC$ computed from generation $Gen_t$ and specific variable costs $varC$ in Eq. \ref{eq:1h}.

In the example, time-series data includes the capacity factor $cf_t$ and demand $dem_t$. Demand is given in power units and is scaled according to the respective step-size by the parameter $\alpha$. At a step-size of 2 hours for instance, $\alpha$ has a value of 2 scaling a demand of 40 GW to 80 GWh. As a result, scaling changes the level of generation and storage variables requiring to scale capacity variables accordingly. Note that the energy capacity of storage is not scaled, because it is already denoted in energy units.
 
If the year is compressed by TSR, energy demand covered in the energy balance does not correspond to actual demand. For example, if the reduced time-series consists of 4,380 hourly steps, all generation variables will only sum to 50\% of actual demand. To correct this, the parameter $\beta$ scales-up all other occurrences of the generation, storage input, or storage output variable outside of the energy balance, storage balance or the capacity constraints. In the stylized example this only affects the computation of variable costs in Eq. \ref{eq:1h}.

\begin{subequations}
\begin{alignat}{4}
\text{min} & \; \; InvC + VarC & & & \label{eq:1a} \\
\text{s.t.} & \; \; Gen_t + St^{out}_t - St^{in}_t & \; \;  = \; \; & dem_t \cdot \alpha & \;  \;  \forall t \in T \label{eq:1b} \\ 
& \; \; St^{size}_{t-1} - St^{out}_t + St^{in}_t & \; \;  = \; \;  & St^{size}_{t} & \;  \; \forall t \in T \label{eq:1c} \\ 
& \; \; cf_t \cdot Capa_{gen} \cdot \alpha & \; \; \geq \; \;  & Gen_t   & \; \; \forall t \in T \label{eq:1d} \\ 
& \; \; Capa_{st} \cdot \alpha  & \; \; \geq \; \; & St^{out}_t + St^{in}_t & \; \; \forall t \in T \label{eq:1e} \\ 
& \; \; Capa_{size} & \; \; \geq \; \; & St^{size}_{t} & \; \; \forall t \in T \label{eq:1f} \\
& \; \; \sum_{\forall i \in I } Capa_{i} \cdot  invC_i & \; \; = \; \; &  InvC  & \; \; \label{eq:1g} \\ 
& \; \; \sum_{\forall t \in T} Gen_t \cdot \beta \cdot  varC & \; \; = \; \; &  VarC  & \; \;  \label{eq:1h} 
\end{alignat}
\end{subequations}

Since $\beta$ compensates any artificial shortening of the year induced by a compressed time-series, the parameter can be defined as the inverse to the share of demand covered by the summed generation variables. This share can again be expressed as the number of time-steps $m$ times the step-size $\alpha$ relative to the total number of hours, which translates into the relation displayed in Eq. \ref{eq:2}. 

\begin{equation} \label{eq:2}
\beta = \left(\frac{m \cdot \alpha}{8760}\right)^{-1} =  \frac{8760}{m \cdot \alpha}
\end{equation}

\begin{figure}
	\centering
		\includegraphics[scale=.4]{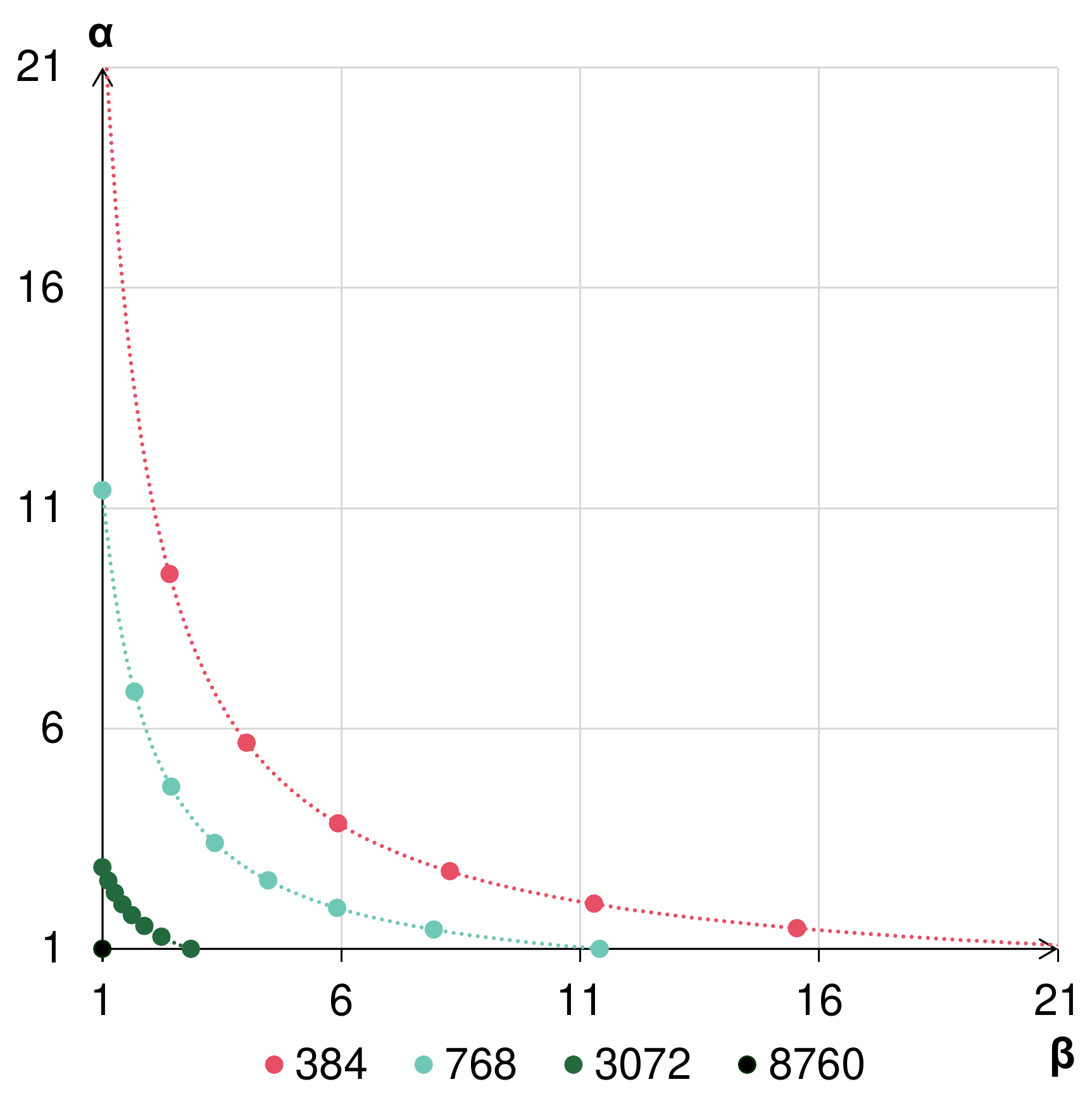}
	\caption{Curves of valid scaling factors}
	\label{fig:3}
\end{figure}

Applying this relation shows that, if a reduced time-series of 384 time-steps is implemented with an hourly step-size implying an $\alpha$ of 1, $\beta$ equals $\scriptstyle \frac{8760}{384}$. If conversely the time-series should be implemented uncompressed, $\beta$ takes a value of 1 and instead the step-size $\alpha$ increases to $\scriptstyle \frac{8760}{384}$. In addition, infinite valid combinations of $\alpha$ and $\beta$ in between these two points exist. The graph in Fig. \ref{fig:3} plots these combinations for different lengths of the reduced time-series $m$ and shows how scaling factors decrease, when temporal resolution increases until $\alpha$ and $\beta$ ultimately converge to one for an hourly resolution.

At first glance, our formalization of implementation methods may appear captious. After all, model results should not change, regardless of whether generation takes a smaller value in the energy balance, but is re-scaled in other places, or takes a larger value in the first place. However, this only holds true in the absence of any inter-temporal effects, like energy storage. If models include storage, scaling step-size or compressing the year imposes a bias---each in a different way. If step-size is increased, demand that is spread across a longer time span is allocated to a single time-step. Therefore, any fluctuations and mismatches of supply and demand within that time-step are neglected and investment into short-term storage to address them is underestimated. On the other hand, only compressing the TSR, but leaving step-size unchanged, neglects seasonal fluctuations. To illustrate this, consider a reduced time-series of 384 steps. Since these steps were chosen to be representative for the entire year, average demand complies with average demand of the full time-series, but total demand is smaller by a factor of $\scriptstyle \frac{8760}{384}$. As a result, short-term storage systems with small energy capacities are capable to shift energy from beginning and middle to the time-series, representing spring and autumn, and effectively operate as seasonal storage.

\section{Derivation of reduced time-series} \label{3}

The other important characteristic of TSR is how to derive the reduced from the full time-series. Depending on whether a derivation method requires or prevents weighting of time-steps, it can only be implemented as grouped periods or as a chronological sequence. If weighting is optional, a method can be used with both.

Table \ref{tab:scenarios} gives an overview of which derivation methods are applied for which implementation method in this paper. The methods were chosen so that a broad range is covered, and the most common approaches are included. The first column specifies the method itself, while the second column lists all included variations. Input to all clustering methods are demand and capacity factors for wind and solar for the six different regions according to the model description in section \ref{4}. All time-series were normalized to be equally weighted by the respective reduction algorithms.

\begin{table}
  \centering
    \begin{tabular}{lcp{1cm}p{0.1cm}p{0.1cm}p{0.1cm}}
          &        & \begin{tabular}[c]{@{}c@{}}\textbf{Grouped}\\\textbf{periods}\end{tabular}  & \multicolumn{3}{c}{\begin{tabular}[c]{@{}c@{}}\textbf{Chronological}\\\textbf{sequence}\end{tabular}} \\
\midrule
\multirow{2}{*}{\textbf{k-Means}} & centroid & \multicolumn{1}{c}{X}     &  \multicolumn{3}{c}{} \\
\cmidrule{2-6}          & (medoid) & \multicolumn{1}{c}{X}     &  \multicolumn{3}{c}{} \\
\midrule
\multirow{2}{*}{\textbf{Hierarchical}} & centroid & \multicolumn{1}{c}{X}     &  \multicolumn{3}{c}{} \\
\cmidrule{2-6}          & (medoid) & \multicolumn{1}{c}{X}     & \multicolumn{3}{c}{} \\
    \midrule
& 10 bins & \multicolumn{1}{c}{X} & \multicolumn{3}{c}{X} \\
\cmidrule{2-6}  \textbf{Poncelet}        & 20 bins & \multicolumn{1}{c}{X}     & \multicolumn{3}{c}{X} \\
\cmidrule{2-6}          & (40 bins) & \multicolumn{1}{c}{X}     & \multicolumn{3}{c}{X} \\
\midrule
\textbf{Gerbaulet} & & \multicolumn{1}{c}{} & \multicolumn{3}{c}{X} \\
\midrule
\textbf{Re-Sampling} & & \multicolumn{1}{c}{} & \multicolumn{3}{c}{X} \\
\end{tabular}%
\caption{Combinations of methods considered for TSR}%
\label{tab:scenarios}%
\end{table}%

In the following, each of the considered reduction methods is briefly introduced. A comprehensive documentation of each method can be found in the cited publications. When deriving reduced time-series grouped into periods, each of these periods represents one day and different lengths are achieved by selecting a different number of days, as it is the norm.

\begin{itemize}
\item \textbf{k-Means}: The method belongs to the group of partitional clustering algorithms that minimize the Euclidean distance between the center of each cluster and its members \citep{Teichgraeber2019a}. In the centroid case, the center is the mean across all members of the cluster. Whereas in the medoid case the center corresponds to the median, the member of the cluster most similar to all others. The number of clusters is pre-set and corresponds to the number of grouped periods meaning in our case the algorithm clusters days based on their respective time-series. The individual weights for each period then correspond to the number of days assigned to each cluster. This weighting prevents implementation as a chronological sequence but facilitates representation of periods that are rare and extreme.
\item \textbf{hierarchical}: The used agglomerative hierarchical clustering algorithm consecutively merges the closest points into clusters until the desired number of clusters is reached. This process is again based on the Euclidean distance and either the centroid or medoid of the existing clusters \citep{Teichgraeber2019a}. Apart from that, the way clusters translate into grouped periods is analogous to the k-Means algorithm. 
\item \textbf{Poncelet}: This method aims to match the duration curves of the reduced and the full time-series \citep{Poncelet2016a,Poncelet2016b}. For this purpose, each duration curve is stepwise linearized with the number of steps named bins. Afterwards, days for the reduced time-series are selected minimizing the difference between the linearized duration curve of the reduced and full time-series. Originally, the method includes different weightings for each selected day. For this paper it was extended with uniform weighting, so results can not only be implemented as grouped periods, but as a chronological sequence as well. Since periods are solely selected and weighted to match the duration curve of the full time-series, representation of extreme periods is limited.
\item \textbf{Gerbaulet}: The method referred to as "Gerbaulet" in this paper combines heuristic and optimization in a 3-step process \citep{Gerbaulet2017}. In the first step, every 25th (or 49th and so forth) of the full time-series is selected. Afterwards, the resulting time-series is smoothed with a moving average to prevent sharp jumps in the time-series. Finally, the resulting time-series is scaled using a non-linear optimization to match minima, maxima and full-load hours of the reduced time-series with the full time-series. As a result, the method preserves the extrema of each time-series, but not how they are correlated, e.g. time-steps with high demand and low capacity factors.
\item \textbf{Re-Sampling}: In case of re-sampling adjacent time-steps of the full time-series are joined together into a single step. Typically, two-, four- or six-hour blocks are used for this purpose and characteristics of the new time-steps are obtained by averaging the original values.
\end{itemize}

In case of the Poncelet method, for the "40 bins" variation final results did not significantly differ from results for the "20 bins" variation, which is they were omitted from the subsequent analysis. The same applies for "medoid" and "centroid" clustering. To apply the k-means and hierarchical clustering methods their implementation in the TimeSeriesClustering.jl package was used \citep{Teichgraeber2019b}. In addition, the package was extended with the Poncelet method for the purpose of this paper. For details see the supplementary material.

Figure \ref{fig:35} provides a comprehensive recapitulation of how reduced time-series are derived from the hourly time-series and implemented into models. In the first step, derivation selects entire days in case of clustering and the Poncelet method, or specific hours in case of the Gerbaulet method. In case of re-sampling adjacent time-steps are joined into new steps. The implementation method then determines how steps are implemented into a model. In principle, any reduced time-series can either be implemented as grouped periods or a chronological sequence, but in practice it is sensible to combine implementation as grouped periods with derivation methods that assign weights to the selected steps. Implementation as grouped periods sorts the selected steps into short periods, typically days, and applies an individual weight to each period. In a chronological sequence all selected steps are put into chronological order and cannot be weighted individually. Therefore, either all steps are weighted with the same factor, or in case of compression, other model elements are corrected to account for the shorter time-series. For instance, if \textit{step 673} from the example has a demand of 50 GW and the reduced time-series has 48 steps, demand for the step within the model can range from 9,125 GWh ($50 \cdot \alpha$ with $ \alpha = {\scriptstyle \frac{8760}{48}}$) in case of full re-scaling to 50 GWh ($\alpha = 1$) in case of full compression.

\begin{figure}
	\centering
		\includegraphics[scale=.5]{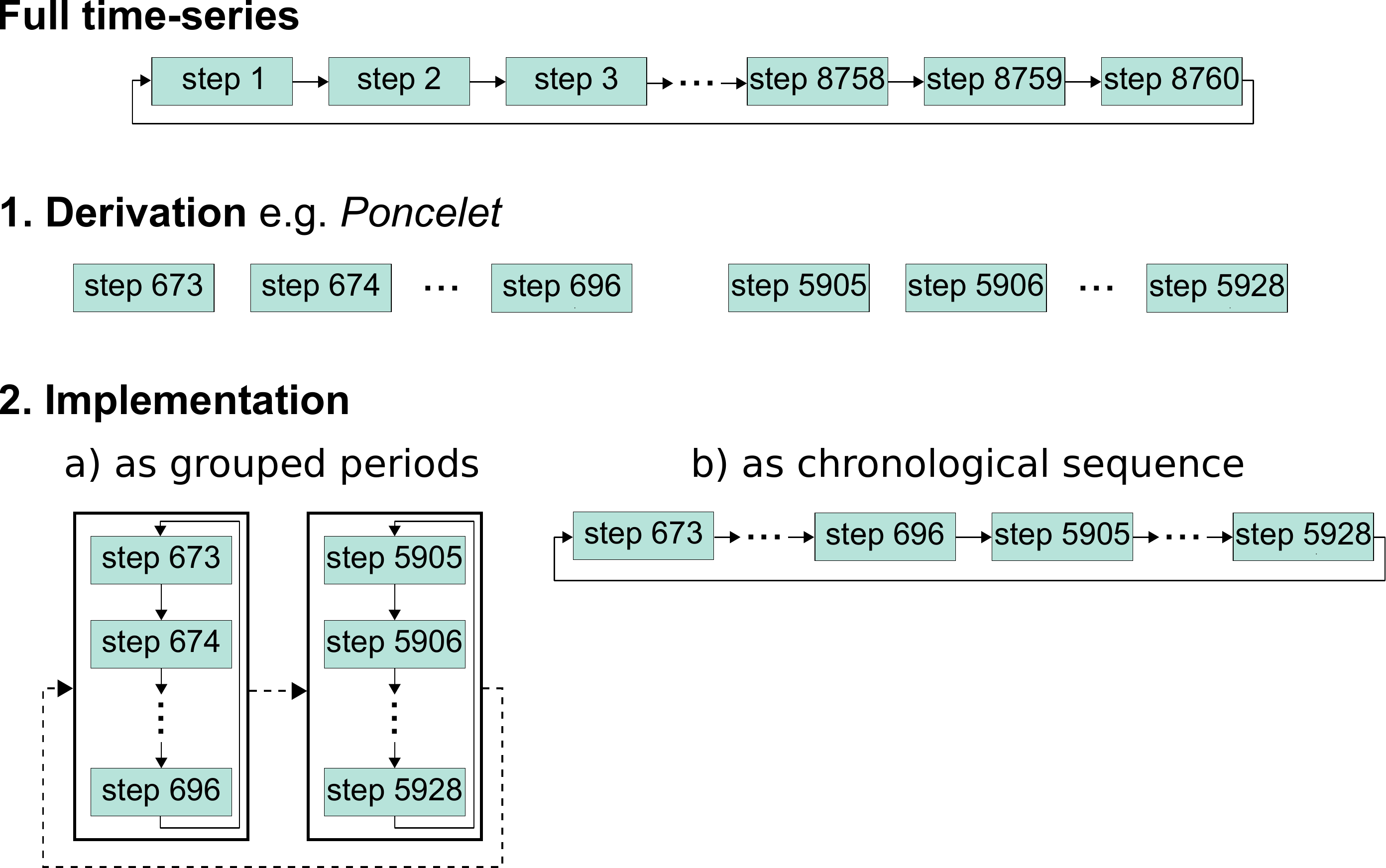}
	\caption{Process to derive and implement reduced time-series}
	\label{fig:35}
\end{figure}
 
\section{Test case for time-series reduction} \label{4}

The various methods for time-series reduction are evaluated based on a stylized capacity expansion model. The model is focused on the power sector due to its high relevance for future energy systems with fully renewable generation. The spatial scope of the model is limited to a single node, resulting in a so-called "island" system. Renewable potential and demand for that node corresponds to data for Germany. Like most large-scale capacity expansion problems, the model is formulated as a linear optimization problem without binary or integer variables, similar to Eqs. \ref{eq:1a} to \ref{eq:1h}.

Demand for electricity includes electricity demand from the heating, industry, and transport sector, based on a scenario where these sectors are fully decarbonized \citep{Hainsch2020}.\footnote{In the respective study full decarbonization is achieved by 2040.} Accordingly, demand totals 956 TWh with conventional applications only accounting for 299 TWh and instead 456 TWh for industrial heating, 91 TWh for residential heating, and 109 TWh for the mobility sector. The applied load profiles also reflect the change in composition of electricity demand. Since decarbonizing other sectors is found to exhaust the available energy potential of biomass, power generation from biomass is excluded. The model does however include the option of shedding load at a cost of 11,000 \euro/MWh \citep{Growitsch2013}. Since estimates of load shedding costs greatly differ across sources, we also tested the model with costs reduced by a factor of 10 to ensure robustness \citep{Praktiknjo2011,Leahy2011}.

\begin{figure}
	\centering
		\includegraphics[scale=.6]{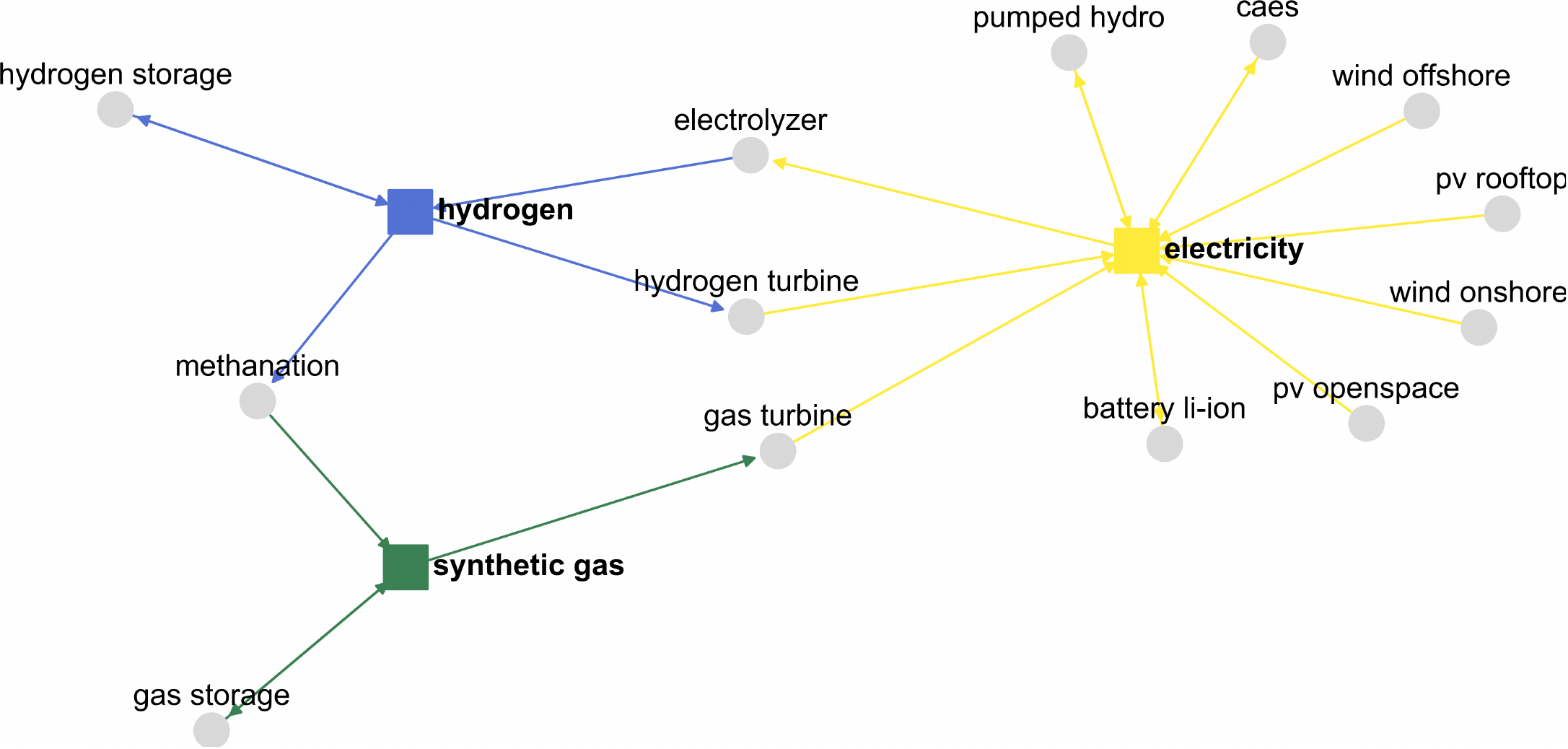}
	\caption{Graph of considered energy carriers and technologies}
	\label{fig:4}
\end{figure}

Figure \ref{fig:4} provides an overview of the technologies and energy carriers considered in the model. In the graph, carriers are symbolized by colored and technologies by gray vertices. Entering edges of technologies refer to input carriers; outgoing edges refer to outputs. 

\begin{figure}
	\centering
	    \includegraphics[scale=0.8]{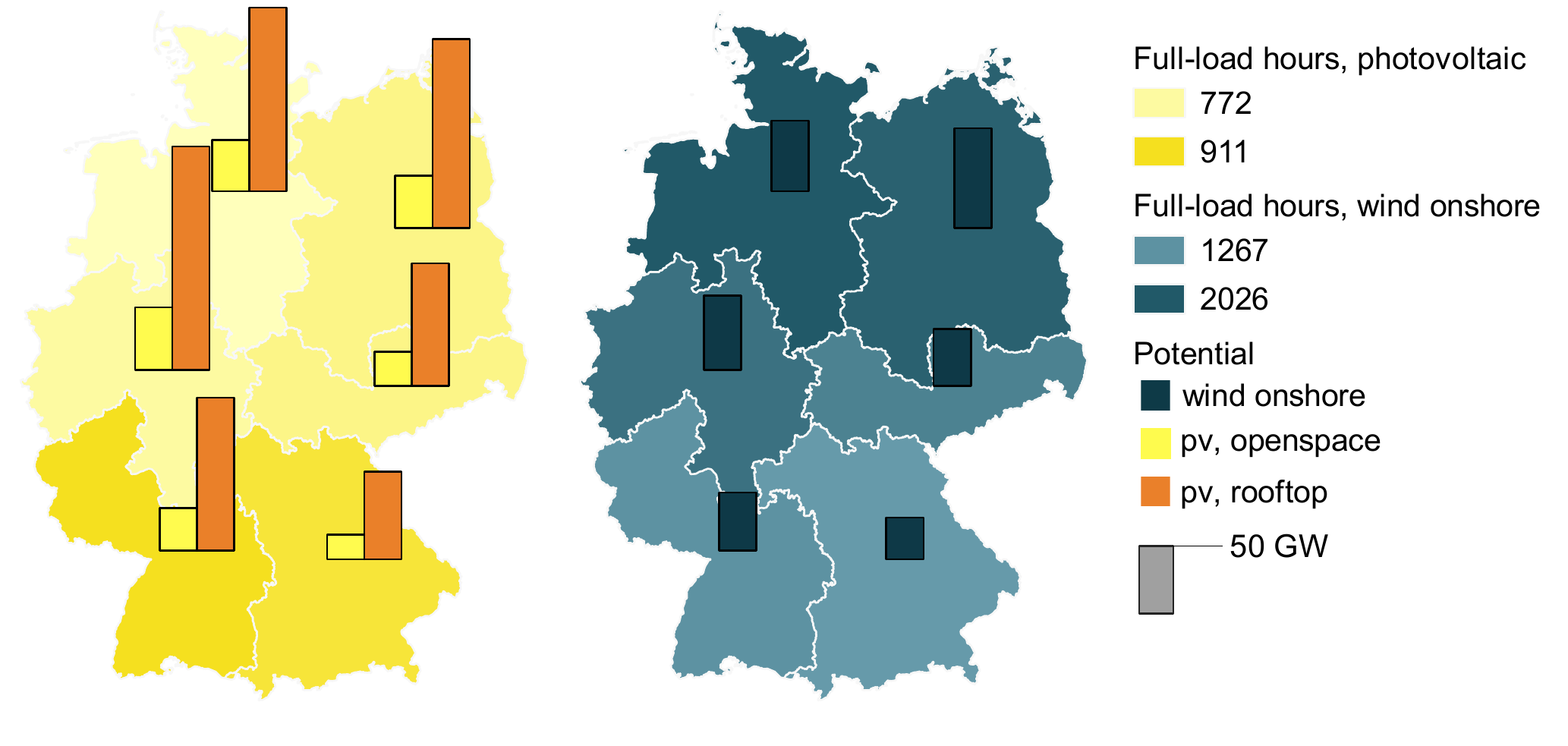}
	\caption{Full-load hours and potential of renewables by region}
	\label{fig:5}
\end{figure}

In accordance with the research question, the model only includes renewable generation technologies like wind, photovoltaic, and run-of-river. Capacity limits and factors for these technologies are differentiated according to the six regions displayed in Fig. \ref{fig:5}. Overall, potential for openspace and rooftop photovoltaic amounts to 198 GW and 707 GW, respectively, for onshore wind a potential of 297 GW is assumed, for offshore wind 84 GW. These limits are based on assumptions used in the H2020 project openEntrance, the time-series of capacity factors are taken from \citet{lkdeu2017}. The analysis separately considers the climatic years 2015 and 2017. In 2015 full-load hours of photovoltaic and wind are close to the long-term average; in 2017 full-load hours are above average for photovoltaic and below for wind.

The included technologies for short-term storage include lithium-ion batteries, pumped-hydro, and compressed air energy storage (CAES). Capacities for pumped-hydro are subject to an upper limit of 7 GW for power capacity and 35 GWh for energy capacity, which matches the capacities installed today. For CAES, power and energy capacity is limited to 6 GW and 24 GWh, respectively \citep{Elsner2015}. For lithium-ion batteries only the ratio of energy to power capacity is restricted and cannot exceed 10. The modeling of seasonal storage is more elaborate and distinguishes the technologies for generation, storage, and re-conversion of synthetic fuels. These include hydrogen generated through electrolysis and used by hydrogen turbines as well as synthetic gas generated through methanation and used by conventional gas turbines.

\begin{figure}
	\centering
		\includegraphics[scale=.25]{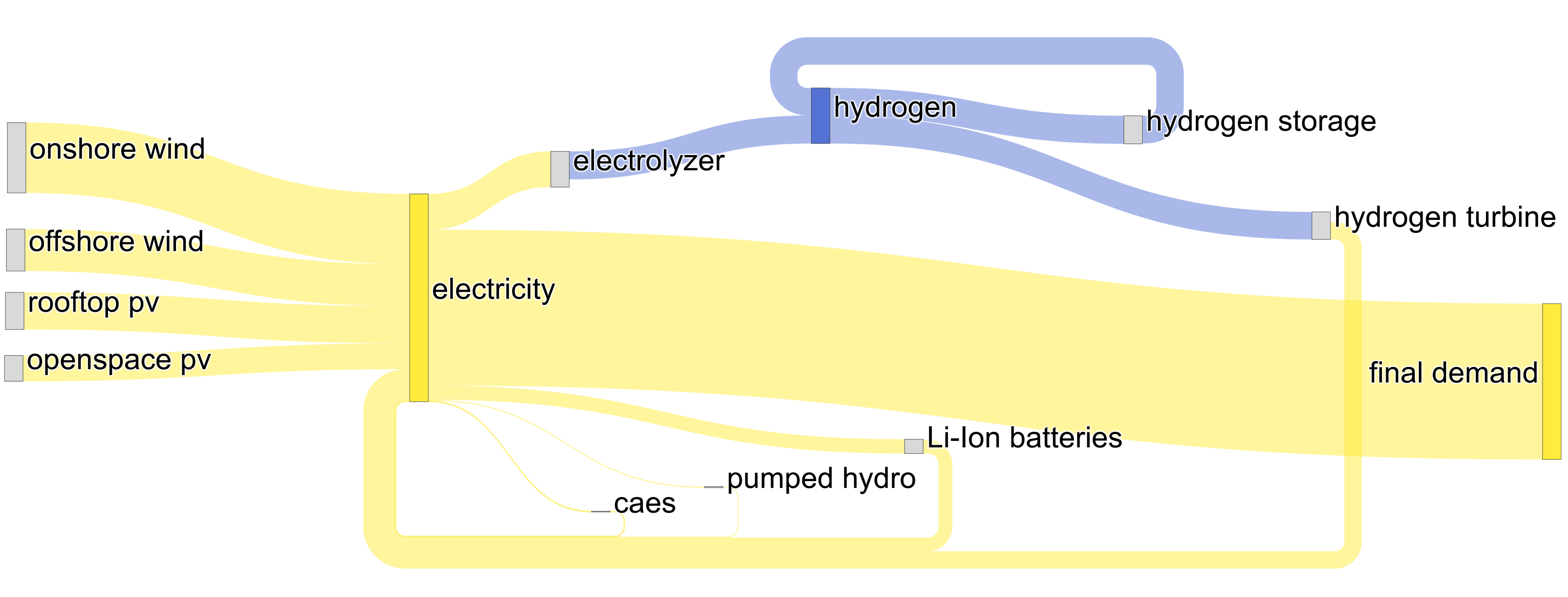}
	\caption{Energy flows when solving with full time-series}
	\label{fig:6}
\end{figure}

The energy flows that result from solving the model with a full time-series are provided by the Sankey diagram in Fig. \ref{fig:6}. Although the entire potential for CAES and pumped-hydro is exploited, short-term storage is still dominated by lithium-ion batteries. Seasonal storage of electricity is achieved by hydrogen, while synthetic methane does not play a role. Load shedding is not used, even in runs with the value of lost load reduced by a factor of 10. Total system costs amount to 61 billion of which 85\% relate to photovoltaic and wind, 9\% to long-term and 6\% to short-term storage.

For the subsequent analysis of TSR, the model will first be solved with a reduced time-series. In a second step, the model is run again, but with a 8,760 hourly time-steps and capacities fixed to the values computed using TSR to evaluate, if demand can be satisfied. Evaluation includes all the combinations of implementation and derivation methods listed in Table \ref{tab:scenarios} and various lengths of the reduced time-series ranging from 24 to 7,680 time-steps.

All model runs implementing reduced time-series as grouped periods were carried of with the Calliope framework, which includes the aforementioned extension for seasonal storage by \citeauthor{Kotzur2018b} \citep{Pfenninger2018}. Since Calliope does not support the different scaling methods for chronological sequences outlined in section \ref{22}, evaluation of chronological sequences was instead performed with the AnyMOD.jl framework \citep{Goeke2020a,Goeke2020b}. To guarantee that observed differences result from the underlying methods and not from the use of different frameworks, we ensured Calliope and AnyMOD.jl return the exact same results when run with a full hourly time-series. The supplementary material provides additional documentation of input data and used modeling tools. 

\section{Results} \label{5}

The results section focuses on how TSR impacts the capability to answer key question of modeling, what are feasible designs for renewable energy systems and what are their respective costs? Consequently, the section is centered on two metrics: First, the share of unmet demand, or loss of load, when running the model at full temporal resolution, but with the capacities obtained using TSR. Second, the deviation of system costs when using a reduced time-series compared to the reference case with full temporal resolution. 

System costs when running at full temporal resolution, but with the capacities obtained using TSR, are not compared, because they are neither informative nor robust results. The only difference to system costs computed with TSR are costs associated with loss of load, that already amount to 80\% percent of total system costs at 1\% of lost load. As a result, system costs when running at full temporal resolution, but with the capacities obtained using TSR, are highly correlated with the loss of load and sensitive to the assumed value of lost load, that greatly differs across sources.

Based on the outlined test case, the first two section analyze TSR implemented as grouped periods or chronological sequences. Subsequently, two sensitivities of the test case are investigated and effects on solve time are discussed.

\subsection{Implementation as grouped periods}

\begin{figure}
	\centering
		\includegraphics[scale=.6]{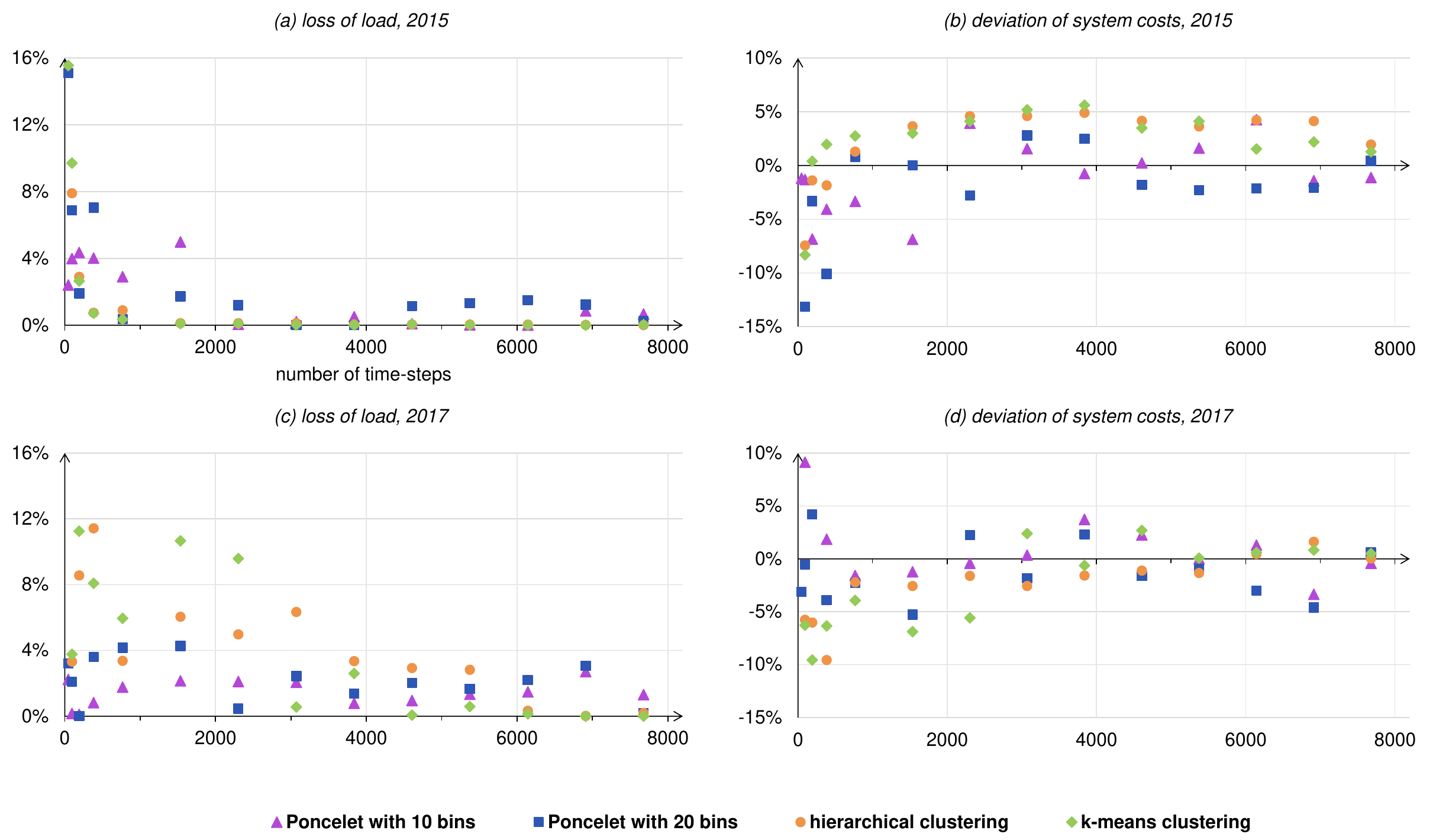}
	\caption{Loss of load and deviations of systems costs for TSR using grouped periods}
	\label{fig:7}
\end{figure}

For all reduction methods implemented as grouped periods, Fig. \ref{fig:7} provides the loss of load on the left and deviation of system costs on the right, both depending on the length of the reduced time-series. Results in the first row relate to 2015; results in the second to 2017. If the reduced time-series is short, all methods exhibit significant lost load of up to 15 \%, that declines as length of the reduced time-series increases. This process is non-monotonic and subject to strong outliers, since the selection procedure of each reduction method introduces some randomness. None of the different methods for creating reduced time-series consistently performs better than the others. Overall, loss of load is much smaller for 2015, the year with average full-load hours, compared to 2017, which has above average full-load hours for photovoltaic.

Deviation of system costs shows a close and negative correlation with the share of lost load, which is plausible considering additional investments can prevent loss of load but increase system costs. When the reduced time-series is comparatively short, a small loss of load is only achieved at the expense of overestimating system costs substantially. Increasing the length of the reduced time-series mitigates this trade-off and has capacities converging towards the optimal design of the system.

\begin{figure}
	\centering
		\includegraphics[scale=.4]{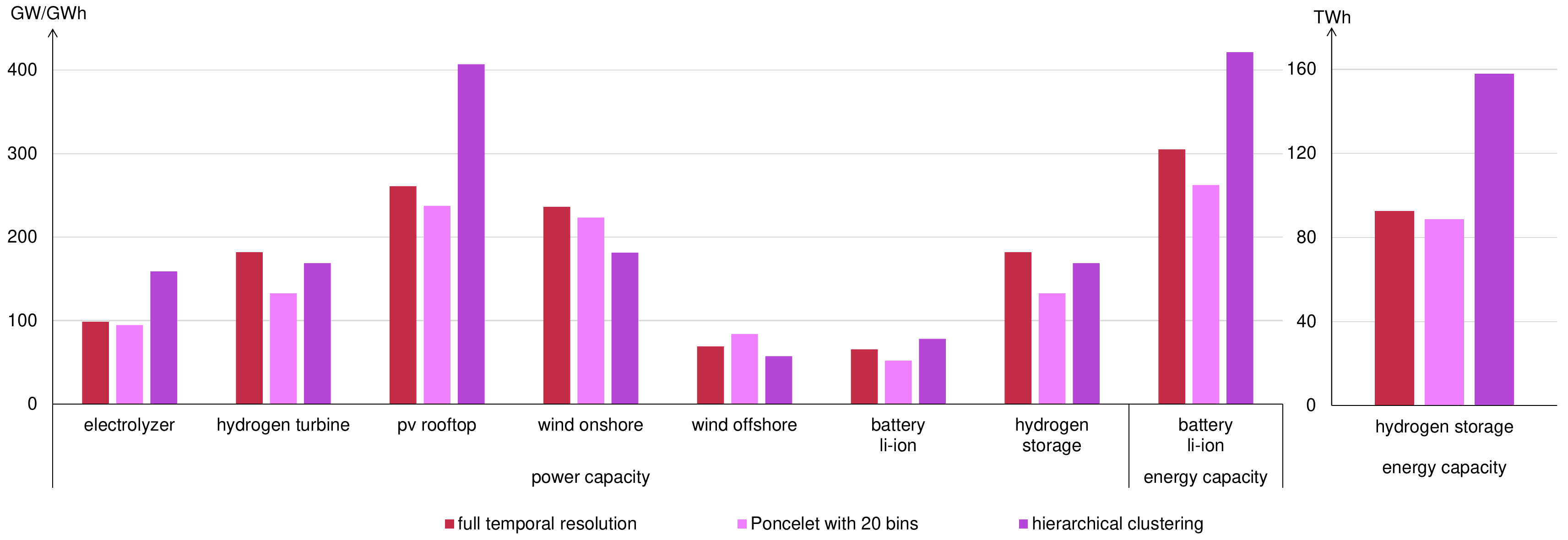}
	\caption{Comparison of installed capacities for different reduction methods}
	\label{fig:8}
\end{figure}

Fig. \ref{fig:8} compares installed power and energy capacities obtained with full resolution against a reduced time-series of 768 time-steps derived using the Poncelet method with 20 bins or hierarchical clustering for the climatic year 2015. Openspace photovoltaic, pumped-hydro and caes are not included, because their technical potential is fully utilized in each case. Other technologies, like methanation or gas turbines, are omitted, because they are not invested in at all. In this example, Poncelet results in 0.8 \% of lost load with unmet demand occurring in 107 hours of the year, compared to 0.9 \% and 331 hours for hierarchical clustering. Poncelet overestimates system costs by 0.7\%; hierarchical clustering by 1.2\%.

The comparison shows that compared to a full time-series and Poncelet, hierarchical clustering finds much greater capacities for rooftop photovoltaic. Since generation from photovoltaic peaks in summer, technologies for seasonal storage, like electrolyzers, hydrogen turbines, and hydrogen storage, serve as a complement to shift generation from summer to winter and are consequently overestimated by hierarchical clustering as well. This relation among errors can be observed in other cases as well, but does not appear more frequently with a certain derivation method. Overall, results revealed no fundamental bias on installed capacities from the respective derivation method.

\subsection{Implementation as a chronological sequence}

\begin{figure}
	\centering
		\includegraphics[scale=.6]{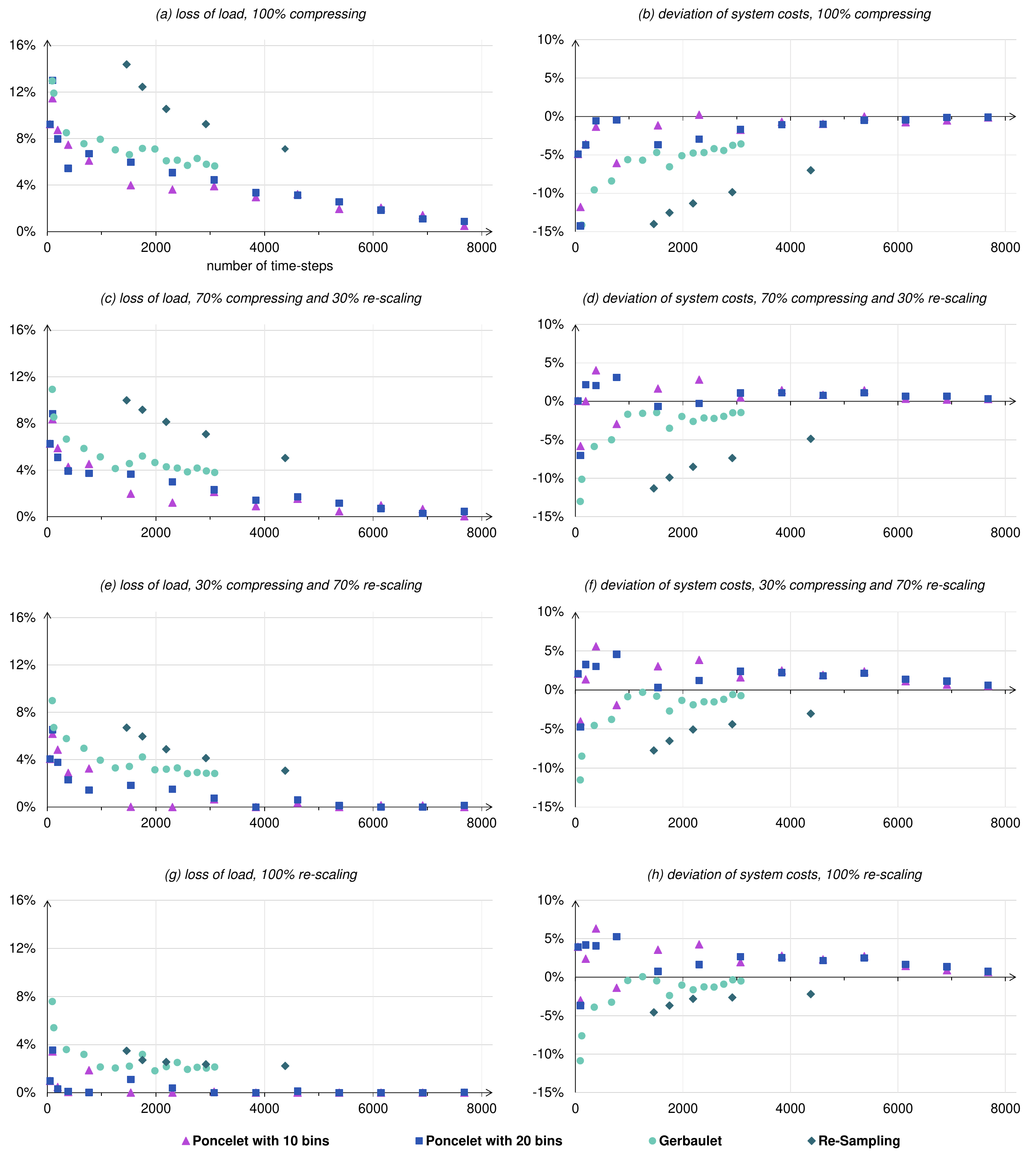}
	\caption{Loss of load and deviations of systems costs for TSR using a chronological sequences and climatic year 2015}
	\label{fig:9}
\end{figure}

\begin{figure}
	\centering
		\includegraphics[scale=.6]{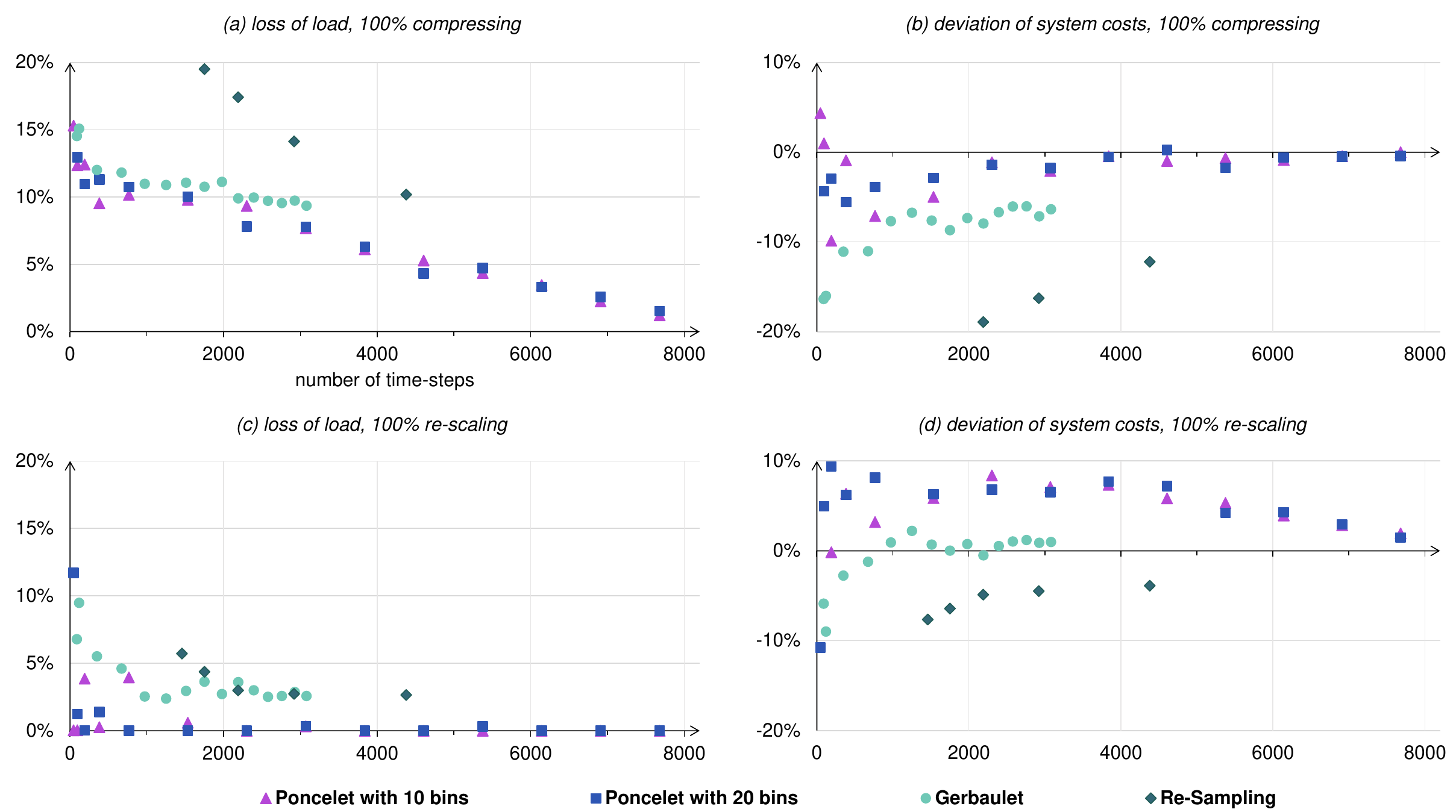}
	\caption{Loss of load and deviations of systems costs for TSR using time-steps and climatic year 2017}
	\label{fig:91}
\end{figure}

Analogously to Fig. \ref{fig:7} for grouped periods, Fig. \ref{fig:9} provides the loss of load and deviation of system costs for chronological sequences in 2015; Fig. \ref{fig:91} provides the same results for 2017. Section \ref{22} introduced two options for implementing chronological sequences, either compressing or or re-scaling the reduced time-series. Results when reduced time-series are compressed are provided in the first row of Fig. \ref{fig:9}. In the following rows the method successively shifts until the forth row finally shows results if time-steps are strictly re-scaled. Since results do not substantially differ for 2017, Fig. \ref{fig:91} only shows results for strictly compressing or re-scaling.

The left column shows the share of lost load when capacities computed with a reduced time-series are tested with a full time-series. Loss of load decreases with increasing length of the reduced time-series, in particular when the time-series is still comparatively short, since more time-steps can capture characteristics of the full time-series more accurately. As observed for grouped periods, the loss of load is substantially smaller for the climatic year 2015 than for 2017. Of all reduction methods, "Poncelet" consistently achieves the smallest loss of load, regardless of how many bins are considered, followed by "Gerbaulet" and lastly Re-Sampling. 

However, the most significant impact on loss of load is not how the full time-series was reduced, but how the chronological sequence was implemented. Loss of load is highest if the reduced time-series is compressed. As implementation shifts towards demand, loss of load strictly decreases and for 2015 ultimately drops below 4\% in all cases but one. System costs are again closely correlated with the share of lost load. High shares imply insufficient investment and a consequential underestimation of system costs. Higher investments in other cases decrease loss of load, but also drive up system costs, eventually even overestimating them.

\begin{figure}
	\centering
		\includegraphics[scale=.45]{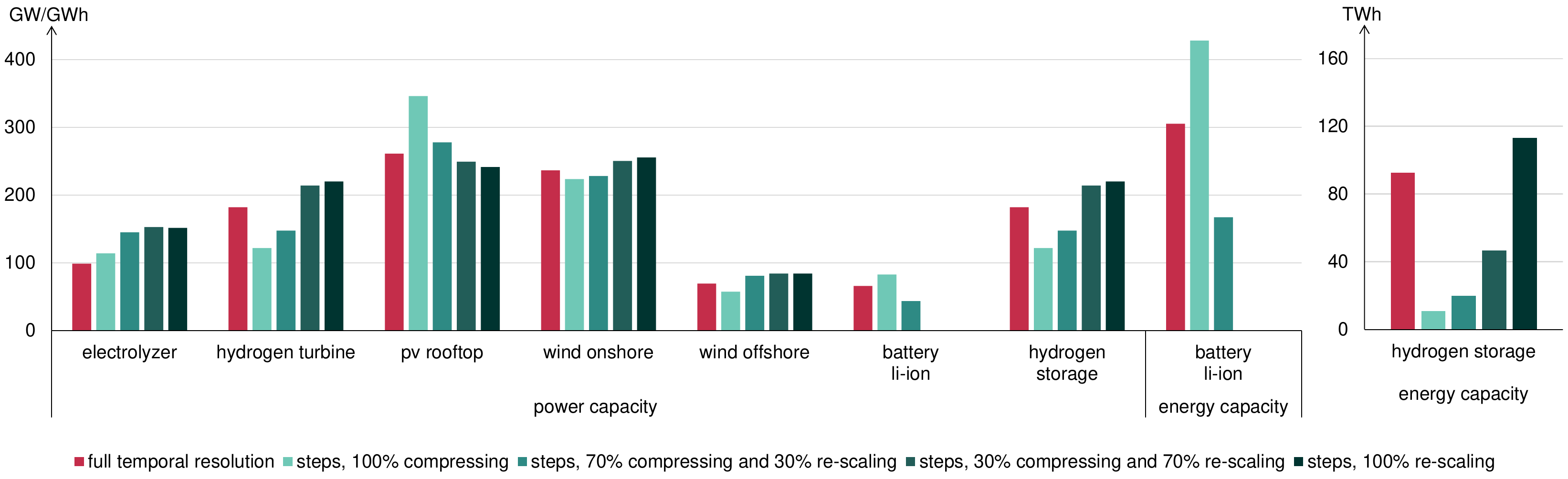}
	\caption{Comparison of installed capacities for climatic year 2015 and different scaling methods}
	\label{fig:10}
\end{figure}

Implementation also has significant impact on installed power and energy capacities, which are compared in Fig. \ref{fig:10} for the Poncelet method with 20 bins and a reduced time-series of 768 time-steps. Analogously to Fig. \ref{fig:8}, technologies with the same investment across all cases are omitted. The graph confirms the bias of scaling proposed in section \ref{22}.

Compressing the reduced time-series overestimates investment into batteries, a technology for short-term storage, but underestimates investment into hydrogen turbines and storage, both technologies for seasonal storage. The effect of scaling on renewable generators depends on how they interact with storage. Since generation from photovoltaic peaks in summer, but demand peaks in winter, at one point additional photovoltaic generation can only be used, if it is stored seasonally. As detailed in section \ref{22}, compressing overstates the ability of short-term storage like batteries to provide seasonal storage and consequently overestimates investment into photovoltaic as well.

Results shift into the opposite direction when implementation shifts from compressing to re-scaling. Now the model accurately reflects the storage technologies' capability of seasonal storage, but neglects short-term fluctuations on a hourly or daily scale. Correspondingly, compressing greatly underestimates investment into battery storage, but approximates capacities for seasonal storage very well. Since seasonal storage is reflected well, investment into renewables also shifts from photovoltaic peaking in summer to wind with a more even seasonal profile.

The bias compressing imposes on short-term storage even remains significant if the reduced time-series is comparatively long and loss of load becomes negligible. When a full-time series is re-sampled into blocks of 4-hours, resulting in a reduced time-series of 91 days, for batteries power capacity is still underestimated by 73\% and energy capacity by 23\%. If sampled into 2-hour blocks, these values decrease to 12\% and 6\%, respectively. 

The discussed results on installed capacities, in particular for storage, also explain the effect scaling has on loss of load. Since compressing overstates batteries' capability of seasonal storage, investment into actual seasonal storage is insufficient and a significant amount of demand cannot be met, mostly during the winter months. Accordingly, if re-scaling is fully applied, in the example 6.6\% of demand are unmet and loss of load occurs in 1126 hours, all of them in first or fourth quarter of the year. The implications of neglecting hourly and daily fluctuations in case of re-scaling are not as severe. Since the installed capacities for seasonal storage are technically also capable to balance short-term fluctuations, lost load is much smaller and only occurs in nine hours totaling 0.001\textperthousand \space of demand.

\subsection{Sensitivities for seasonal storage and sector integration}

The test case introduced in section \ref{4} differs from today's system in many ways. Not only does the test case exclusively include renewable generation technologies but is also distinguished by the role of seasonal storage and change of demand. Besides sector integration doubling total demand, new applications for electricity, in particular electric heating, greatly affect the profile of demand, seasonality becomes more pronounced and peak loads increase. These differences raise the question if the preceding results are only caused by intermittent renewables or also must be attributed to storage and demand.

\begin{figure}
	\centering
		\includegraphics[scale=.6]{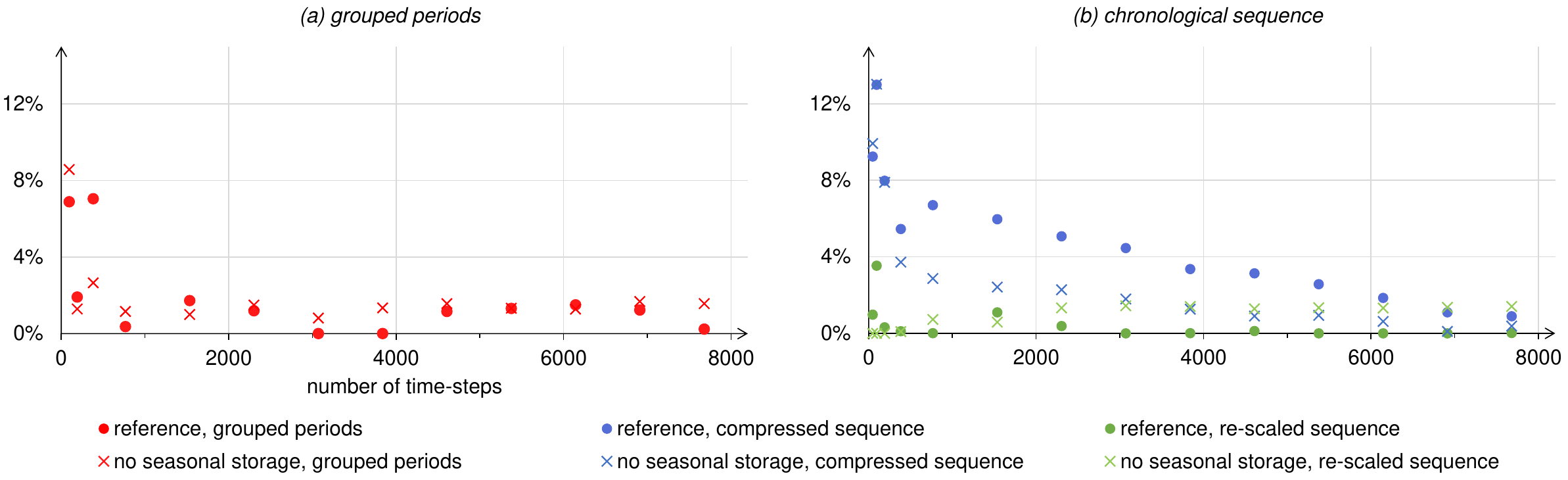}
	\caption{Loss of load sensitivity for seasonal storage}
	\label{fig:11}
\end{figure}

Fig. \ref{fig:11} shows how loss of load for the reference case and climatic year 2015 compares against a sensitivity without any technologies for seasonal storage. In the figure, length of the reduced time-series and implementation method vary, but all cases apply the Poncelet method with 20 bins. Note that excluding seasonal storage only serves the purpose of analysis and is not a practical scenario since system costs double and battery capacity increase by a factor of eight. 

Again results exhibit a strong random variation, but at least for compressed chronological sequences the share of lost load decreases, if seasonal storage is omitted from consideration. This indicates that for this method the error from TSR is not only caused by the intermittency of renewables, but also by difficulties to represent the operation of seasonal storage, which is plausible considering the method's characteristics.

\begin{figure}
	\centering
		\includegraphics[scale=.6]{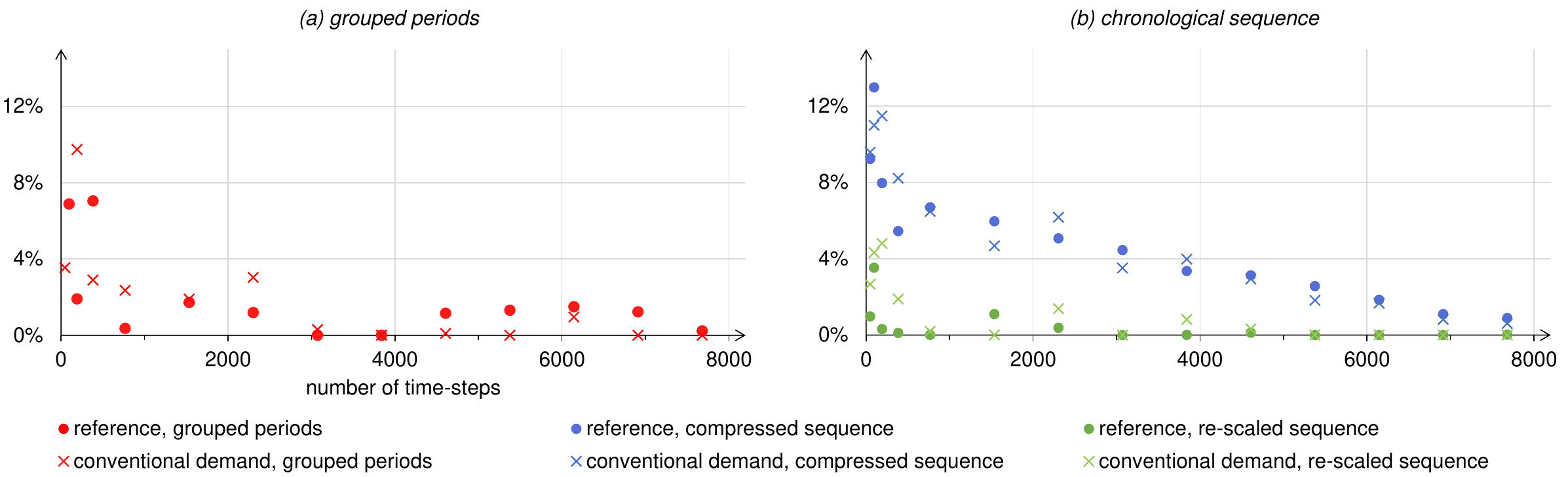}
	\caption{Loss of load sensitivity for demand}
	\label{fig:12}
\end{figure}

Analogously to Fig. \ref{fig:11}, Fig. \ref{fig:12} compares the reference case against a sensitivity on demand, that does not account for sector integration and uses conventional demand data from Germany in 2015 instead. In this case, results do not significantly improve or worsen for any of the implementation methods suggesting that how sector integration changes demand does not affect TSR.

\subsection{Impact on computation time}

\begin{figure}
	\centering
		\includegraphics[scale=.45]{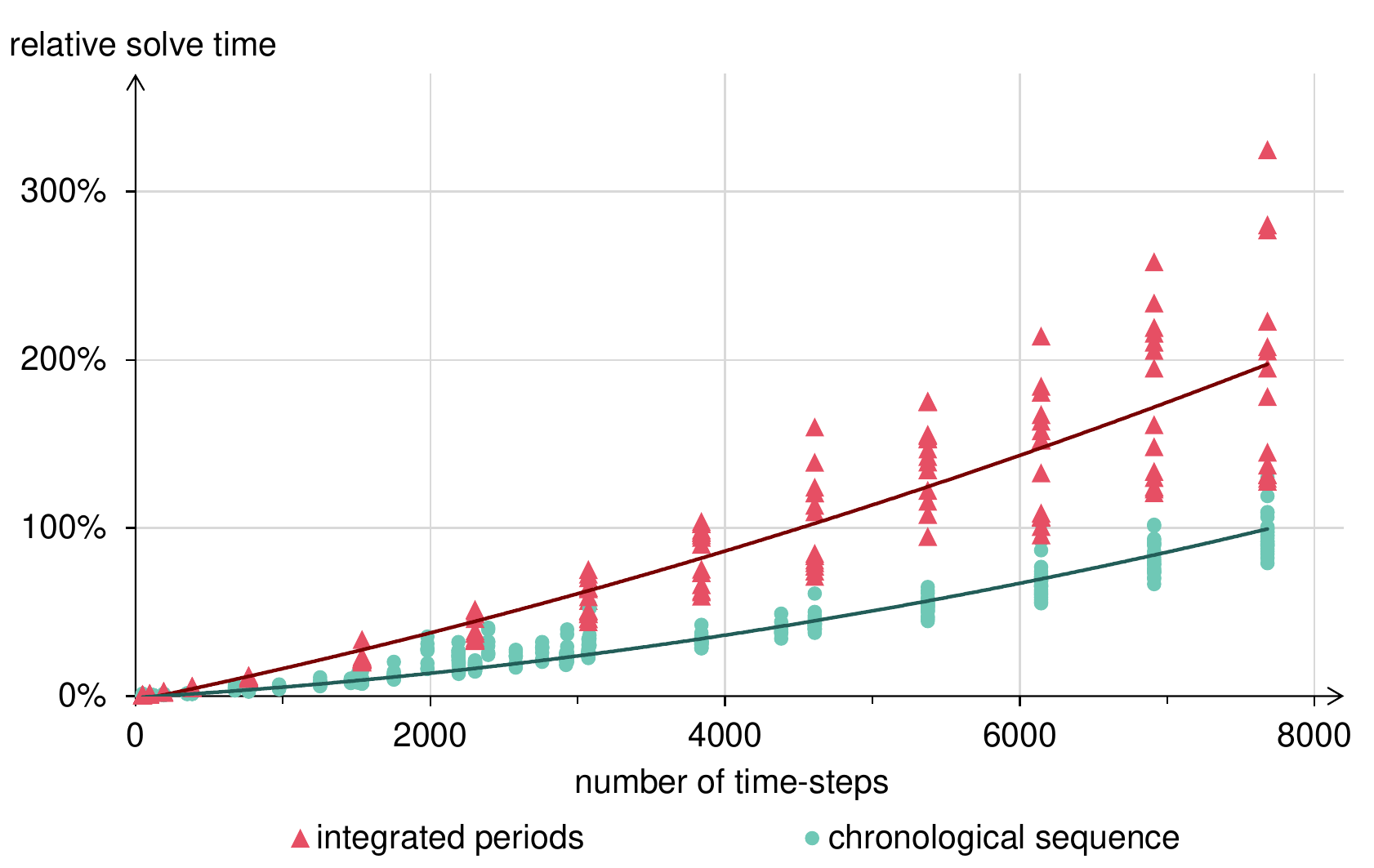}
	\caption{Solve time relative to full temporal resolution}
	\label{fig:13}
\end{figure}

Since the original motivation for TSR is to facilitate solving large models, Fig. \ref{fig:13} analyses the impact the implementation method and the number of time-steps have on solve time. All solve times are denoted relative to the solve time of the model at full temporal resolution to adjust for the effect of the respective modeling framework.\footnote{AnyMOD.jl solved the full model in 65 seconds; Calliope in 330 seconds, both producing the exact same results.} The reported times only include optimizer time. All models were solved on the same cluster with Gurobi using the Barrier method and 2 threads.\footnote{The Gurobi parameters BarOrder and NumericFocus were set to 1 and 2, respectively.} Since solve times are subject to random variations when solving models with the same number of time-steps but different data, quadratic regression curves were added to the figure to visualize the general trend.

As expected, the reduction of solve time strongly depends on the number of time-steps used. At numbers around 3,000 that were found earlier to considerably reduce loss of load, TSR still reduces solve-time significantly. For chronological sequences, solve times converge towards the solve time with full temporal resolution, which is plausible considering chronological sequences are implemented analogously to a full time-series. Solve time for grouped periods increases faster and can even exceed the solve time with full temporal resolution. This effect is explained by the inter-period variables and constraints added to the model increasing its complexity. 

\section{Conclusion and outlook} \label{6}

The results show that TSR should be applied with caution when modeling renewable energy systems. Besides intermittency of renewables, dependency on seasonal storage adversely affects the accuracy of TSR. As suggested by former research, we found the accuracy of TSR to increase with the length of the reduced time-series. 

Implementation of reduced time-series as grouped periods did not consistently achieve small shares of lost load. Furthermore, no generally advantageous method for creating reduced time-series nor any fundamental bias on installed capacities was identified. Compared to chronological sequences, grouped periods required more time so solve for the same number of time-steps, presumably due to variables and constraints added to implement seasonal storage.

For implementation as chronological sequences, results highly depend on how reduced time-series are adjusted to achieve consistency with the full time-series. If the reduced time-series is not re-scaled, results show a bias towards short-term storage and considerable loss of load. If the reduced time-series is re-scaled on the other hand, results are instead biased towards long-term storage overestimating system costs but achieving small shares of lost load. These results are likely to differ for regions with less pronounced seasonal and greater short-term fluctuations. Regarding the derivation of reduced time-series, the Poncelet method performed favorable with chronological sequences.

Results for macro-energy systems in this paper cannot be transferred to models of small-scale systems. In this case, flexible grid supply mitigates the problem of adequacy and integer constraints cause a steeper increase of computational complexity, rendering TSR more favorable overall. 

Our benchmark for TSR in this paper is a single climatic year at an hourly resolution, which is not exhaustive. The impact the respective climatic year had on our results, which is consistent with other studies, suggests to extend the scope of planning to multiple climatic years \citep{Bloomfield2015,Hilbers2019}. In addition, adequacy of hourly resolutions can be called into question, since it is based on assuming sub-hourly fluctuations balance out across sufficiently large areas \citep{Deane2014,Brown2018b}. This is in particular questionable, if models further increase spatial resolution to represent renewables more accurately, as frequently proposed \citep{Frysztacki2021,mart}. Lastly, spatial and temporal detail must be weighed against methodological simplifications to reduce complexity and keep models linear. For instance, the capacity expansion model applied in this paper assumes that time-series are perfectly predictable and neglects unit commitment or operational constraints of thermal power plants \citep{Seljom2015}. So, in conclusion, necessary detail beyond out test case adds to the deficiencies of TSR identified in this paper and suggests further efforts to reduce complexity of large-scale capacity expansion models. 

On the one hand, existing methods for TSR can be further improved. One approach is to identify extreme situations in the full time-series where adequacy is threatened and add them to the reduced time-series. However, identifying such situations with low supply and high demand over a prolonged period of time is non-trivial, because both supply and demand again depend on investment decisions by the model. Against this background, \citet{Teichgraeber2020} introduce a method that iteratively adjusts the extreme periods in the reduced time-series by performing a feasibility check with the full time-series. Although their work is focused on small-scale systems, similar approaches could be adopted for macro-energy systems as well. Another approach is to vary temporal resolution within the model and only apply high detail where it is necessary. \citet{Renaldi2017} introduce such a method for small-scale systems; \citet{Goeke2020a} develops a similar approach for macro-energy systems.

On the other hand, complexity can be reduced by partitioning problems into smaller parts. Therefore, one approach is to couple models with different scopes and resolutions instead of using one comprehensive but highly complex model to get a broader picture of the energy system. In this case, typically results of long-term planning models are evaluated with more detailed operational models \citep{Antenucci2019,Collins2017,Pavicevic2020}. A downside of this approach is the limited capability to feed information from the operational model, like hours with unmet demand, back to the planning model. Alternatively to coupling different models, complex models can be decomposed into smaller parts, typically relating to planning or operation as well, and then solved faster with advanced solution methods. With the exception of \citet{Sepulveda2020}, such methods have not yet been adopted for macro-energy systems and are focused on small-scale applications \citep{Yokoyama2015,Bahl2018,Baumgartner2020}.

\section*{Acknowledgements}

The authors thank Christian von Hirschhausen and Dylan Manning for valuable comments on the work and Bryn Pickering for his support with the Calliope framework.

The research leading to these results has received funding from the European Union's Horizon 2020 research and innovation program under grant agreement No 773406 and from the German Federal Ministry for Economic Affairs and Energy under the project acronym LKD-EU (FKZ 03ET4028C).

\section*{Supplementary material}

Three different open-source tools were applied for the research in this paper. These include version 0.1.6 of the modeling framework AnyMOD.jl (\url{https://github.com/leonardgoeke/AnyMOD.jl}), version 0.6.5 of of the modeling framework Calliope (\url{https://github.com/calliope-project/calliope}) and an extension to version 0.5.3 of TimeSeriesClustering.jl (\url{https://github.com/leonardgoeke/TimeSeriesClustering.jl/tree/dev}) specifically created for this paper.

All scripts and data files for creating the reduced time-series and evaluating them with a capacity expansion model are available on Zenodo (\url{https://doi.org/10.5281/zenodo.4992922}). The upload also includes additional information on the input parameters used.

\appendix

\numberwithin{equation}{section}
\makeatletter 
\newcommand{\section@cntformat}{Appendix \thesection:\ }
\makeatother

\setcounter{equation}{0}

\bibliographystyle{model3-num-names.bst}

\bibliography{cas-refs}
\end{document}